# Sputtered terbium iron garnet films with perpendicular magnetic anisotropy for spintronic applications

S. Damerio[1, a)] and C. O. Avci[1]
*Institut de Ciència de Materials de Barcelona (ICMAB-CSIC), Campus de la UAB, 08193 Bellaterra, Spain*

(*Electronic mail: sdamerio@icmab.es, cavci@icmab.es.)

(Dated: 7 February 2023)

We report the structural, magnetic, and interfacial spin transport properties of epitaxial terbium iron garnet (TbIG) ultrathin films deposited by magnetron sputtering. High crystallinity was achieved by growing the films on gadolinium gallium garnet (GGG) substrates either at high temperature, or at room temperature followed by thermal annealing, above 750 °C in both cases. The films display large perpendicular magnetic anisotropy (PMA) induced by compressive strain, and tunable structural and magnetic properties through growth conditions or the substrate lattice parameter choice. The ferrimagnetic compensation temperature ($T_M$) of selected TbIG films was measured through temperature-dependent anomalous Hall effect (AHE) in Pt/TbIG heterostructures. In the studied films, $T_M$ was found to be between 190-225 K, i.e., approximately 25-60 K lower than the bulk value, which is attributed to the combined action of Tb deficiency and oxygen vacancies in the garnet lattice evidenced by x-ray photoelectron spectroscopy measurements. Sputtered TbIG ultrathin films with large PMA and highly tunable properties reported here can provide a suitable material platform for a wide range of spintronic experiments and device applications.

## I. INTRODUCTION

Spintronics, the concept of using electrons' spin as an active variable in electronic circuits, has evolved into a broad and interdisciplinary research field at the intersection of physics, materials science and nanotechnology. Ferrimagnetic insulators have lately attracted significant attention in the spintronics community thanks to their suitable spin transport and magnetic properties.[1,2] Among them, rare-earth iron garnets (REIGs) are particularly interesting thanks to their well-understood and tunable properties through composition, strain, stoichiometry and temperature.[3,4] The use of REIGs in spintronics has led to various pioneering achievements and fundamental discoveries during the past decade, such as the longitudinal spin Seebeck effect,[5] spin Hall magnetoresistance (SMR),[6] long-range spin transport,[7] efficient current-induced magnetization switching,[8] domain wall[9] and skyrmion.[10] motion Low damping, large magnon density, high interfacial spin transparency, and the possibility of obtaining perpendicular magnetic anisotropy (PMA) are among the most attractive properties of REIGs, which facilitated some of the above achievements. Obtaining high-quality ultrathin REIG films with tunable properties is thus an important challenge for spintronic research and future related technologies.

Earlier REIGs films have been fabricated by liquid phase epitaxy.[11,12] This method yields high-quality films with several micrometers of thickness useful for the fundamental understanding of their properties but unsuitable for most spintronic experiments relying on current-driven magnetization control.[13] More recently, pulsed laser deposition (PLD) has emerged as a convenient method to grow nanometer-thick REIG films with bulk-like properties.[14,15] Among REIGs, yttrium iron garnet ($Y_3Fe_5O_{12}$, YIG) has been ubiquitous for a long time in insulating spintronics, mainly due to its record low damping, allowing efficient magnonic signal generation and transmission.[16,17] Lately, however, the focus has started to shift towards REIGs with magnetic rare-earth elements such as $Tm_3Fe_5O_{12}$ (TmIG), $Eu_3Fe_5O_{12}$ (EuIG), $Tb_3Fe_5O_{12}$ (TbIG), etc.[18–23] In these films, it is possible to obtain strain-induced PMA due to the lattice mismatch with $Gd_3Ga_5O_{12}$ (GGG) or similar substrates.[24–26] Moreover, coupled lattice dynamics, together with magnetic and angular compensation in ferrimagnets, allow high-speed and efficient magnetization manipulation by means of magnetic fields[27] and spin-orbit torques.[28–30] In this context, REIGs with high magnetic compensation temperatures ($T_M$) such as TbIG ($T_M =248$ K) and GdIG ($T_M =286$ K) are particularly attractive for room-temperature ultrafast spintronics. Lately there is a growing interest in preparing REIG films by techniques other than PLD, such as magnetron sputtering, which is more desirable for large-area deposition and complementary-metal-oxide-semiconductor production lines of the microelectronic industry.

In this work, we report the deposition and comprehensive characterization of TbIG thin films obtained by radio-frequency (RF) magnetron sputtering. Films in the 20-30 nm thickness range were grown under compressive strain on GGG and scandium-substituted GGG (SGGG) substrates and thus exhibit PMA. The magnetic properties were found to be highly sensitive to the thermal annealing conditions, target-to-substrate (T-S) distance, and substrate choice. In particular, we found that the growth on a heated substrate leads to enhanced magnetic and structural properties compared to crystallizing the films by annealing subsequent to room temperature growth. We also found a deficiency of $Tb^{3+}$ ions in the films when the T-S distance is large, which can be partially corrected by growing closer to the target. Moreover, we measured the SMR-induced anomalous Hall ef-

a)Institut de Ciència de Materials de Barcelona (ICMAB-CSIC), Campus de la UAB, 08193 Bellaterra, Spain

fect (AHE) in TbIG/Pt bilayers and found that, in magnetically and structurally optimized films, the signal is comparable to literature values, evidencing efficient interfacial spin transport. $T_M$ of the selected films have been characterized through temperature-dependent AHE measurements, yielding $T_M \sim 200$ K, i.e., $\sim 50$ K lower than in bulk TbIG, independently of the Tb:Fe ratio of the films. The films grown and characterized in this study may provide a suitable platform for sputter-grown magnetic insulators with tunable properties for spintronic research and device applications.

## II. EXPERIMENTAL DETAILS

We grew TbIG thin films by RF-magnetron sputtering from a stoichiometric target. We systematically studied the effect of: *i*) temperature, *ii*) T-S distance, and *iii*) substrate lattice parameter, while keeping the remaining deposition parameters constant. Prior to depositions, the sputtering chamber had a base pressure below $5 \cdot 10^{-8}$ mtorr and growths were conducted at 3 mtorr in a mixture of Ar and $O_2$ with a ratio of 30:2. The deposition rate was $\sim 0.4$ nm/min at the applied power of 150 W in the standard T-S distance of 150 mm.

To test the effect of annealing on the film quality and properties, we prepared two sets of samples. In the first set, four separate TbIG films were grown on the standard GGG (111) substrates at room temperature (RT). The first three samples were annealed in-situ for 1 h in 50 mtorr of $O_2$ at 700 °C, 750 °C and 800 °C. The fourth sample was annealed at 900 °C ex-situ in a tubular furnace in a constant oxygen flow of 0.5 L/h, as this temperature is beyond the reach of our sputtering chamber. The second set consists of two samples deposited on GGG (111) substrates kept at 750 °C and 800 °C during growth. To test the effect of the T-S distance on the stoichiometry of the films, we varied this parameter systematically between 158 (+8) mm and 138 (-12) mm (numbers in brackets indicate the deviation from the standard T-S distance of 150 mm), while maintaining the growth temperature at 800 °C and all the other parameters constant. Finally, we also grew TbIG on substituted-GGG (111) substrates with lattice parameters $a = 12.48$ Å (denoted as SGGG-48) and $a = 12.50$ Å (denoted as SGGG-50) at 750 °C. All the samples grown at high temperatures were cooled down in a 50 mtorr of oxygen atmosphere at a rate of -0.5 °C/s.

The crystal structure was investigated utilizing x-ray diffraction (XRD) in a Bragg-Brentano geometry and Cu K-alpha radiation ($\lambda$=1.5406 Å). $2\theta - \omega$ scans were collected around the film and substrate (444) peaks, appearing approximately at 51°. The film thicknesses were extracted from the fit of x-ray reflectivity (XRR) measurements. The surface quality and roughness were assessed using atomic force microscopy (AFM) with a constant amplitude dynamic mode setting.

The chemical composition was studied by means of x-ray photoelectron spectroscopy (XPS). The spectra were collected at room temperature in ultrahigh vacuum ($5 \cdot 10^{-8}$ mtorr) using monochromatic Al K-alpha radiation (1486.74 eV). The analysis of the XPS data was performed with the CasaXPS software. For the quantification of the Tb:Fe ratio we used the Tb $3d_{5/2}$ and Fe $2p_{3/2}$ peaks with relative scattering factors of 49.42 and 10.82, respectively.

The magnetic hysteresis and anisotropy axis of continuous films were examined using a home-built magneto-optic Kerr effect (MOKE) setup in a polar geometry, sensitive to the out-of-plane magnetization direction. In addition, the in-plane and out-of-plane magnetization (*M*) vs. field (*H*) loops were collected at RT in a superconducting quantum interference device (SQUID) magnetometer.

To study the interfacial spin transport properties, we fabricated Pt Hall bar devices on continuous TbIG films with the current injection track length of 30 μm and width of 7.5 μm by means of standard optical lithography and lift-off technique. First, negative photoresist was spin-coated onto the films and cured for 1 min at 75 °C. Then, the desired pattern was exposed to UV light using a micro-writer and developed for 1 min at RT. Finally, 4 nm of Pt were deposited by DC sputtering at room-temperature with 50 W power in 3 mtorr of Ar, and the remaining resist was removed with acetone.

For the electrical measurements, the Hall bars were wire bonded and mounted on a motorized stage allowing for in-plane ($\phi$) and out-of-plane ($\theta$) rotation, and placed in an electromagnet, producing fields of up to 2 T. The experiments were performed at RT using an a.c. current (*I*) of the amplitude of 0.5-7 mA (r.m.s.) in the Pt layer, and frequency $\omega$=1092 Hz. The first Harmonic Hall voltage ($V_H$) was measured as a function of the applied magnetic field, and the corresponding Hall resistance ($R_H$) was calculated as the ratio between $V_H$ and *I*. To measure the temperature dependence of the AHE, we performed similar measurements inside a physical property measurement system (PPMS) for selected films in the temperature range of 50 to 300 K and a swept field in the range of ±1T.

## III. RESULTS AND DISCUSSION

### A. Structural characterization

The XRD and XRR patterns of the sputtered TbIG films are shown in Fig. 1. In Fig. 1(a)-(b), we study the effect of post-growth annealing on the films grown at RT (denoted as GRT). We observe that the GRT films are amorphous and have a thickness of $\sim 30$ nm. Annealing at 700 °C (A700) is insufficient to crystallize the sample, but the increased period and reduced amplitude of the XRR oscillations indicate that the thickness starts to reduce and the surface roughness increases. Upon annealing the films at 750 °C and above, we observe clear XRD peaks around 50° corresponding to TbIG (444), indicating crystalline and epitaxial films. We note that the (444) peaks appear towards smaller angles compared to the bulk value ($d_{444}$=1.7984 Å$^{-1}$) corresponding to the black line in Fig. 1(a), due to the out-of-plane elongation of the unit cell induced by the in-plane compressive strain. This strain state is the origin of the PMA in TbIG films, as discussed later.

In Fig. 1(c)-(d), we report the XRD and XRR data of the films grown on the heated GGG substrate at 800 °C with vary-ing T-S distance. All films show lower spreading of the dispersion of the (444) lattice spacing (based on the full width at half maximum values of the peaks in the $2\theta - \omega$ scans) and multiple thickness fringes, evidencing higher crystal quality than the post-annealed samples (namely, A750, A800 and A900). Moreover, the (444) peaks are observed at smaller angles compared to post-annealed samples demonstrating a higher strain, which is likely to increase PMA in these films. Interestingly, varying T-S distance also leads to a gradual change in the out-of-plane lattice parameter, which is attributed to the distance-dependent systematic variation in the Tb content. We find that the lowest T-S distance results in the highest strain state, which coincides with better stoichiometry according to the XPS data, as discussed later.

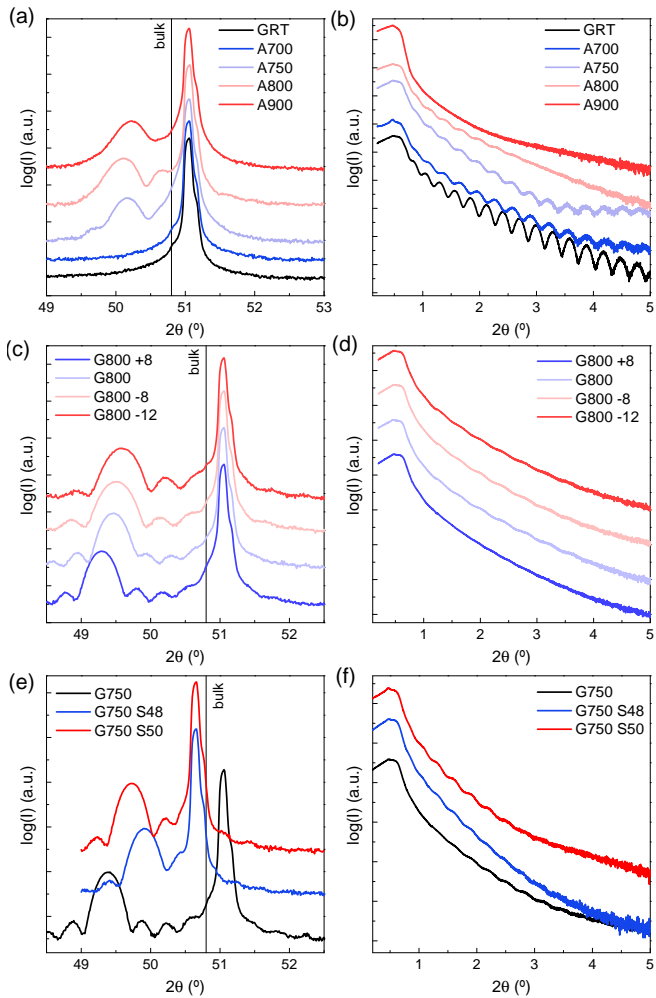

FIG. 1. Structural characterization by X-ray diffraction (XRD). $2\theta - \omega$ scans and reflectivity of 22-29 nm thick TbIG films: (a,b) grown at RT and annealed at increasing temperature and (c,d) grown at 800 °C at varying T-S distance and (e,f) grown at 750 °C on different GGG substrates. The bulk (444) peak position is marked with a black line.

In Fig. 1(e)-(f), we report the XRD and XRR data of the films deposited on heated GGG and SGGG substrates at 750 °C. Here, we confirm that the growth on GGG at this tem-perature results in a film with similar quality with respect to the growth at 800 °C. To our surprise, the films grown on SGGG substrates are also crystalline and display larger $d_{444}$ than bulk. This is unexpected, given that the substrate's lattice parameter is larger than that of bulk TbIG ($a = 12.46$ Å) and thus should not induce any in-plane compressive strain (see Supplementary Material). We tentatively explain this behavior by an expansion of the garnet lattice due to the presence of cation vacancies.[31] Moreover, in this case, the strain does not decrease proportionally to the substrate's lattice parameter, but a smaller out-of-plane strain ($\sim$1.5%) is found on SGGG-48 than on SGGG-50 ($\sim$1.9%). The parameters extracted from the XRD measurements are reported in Table I.

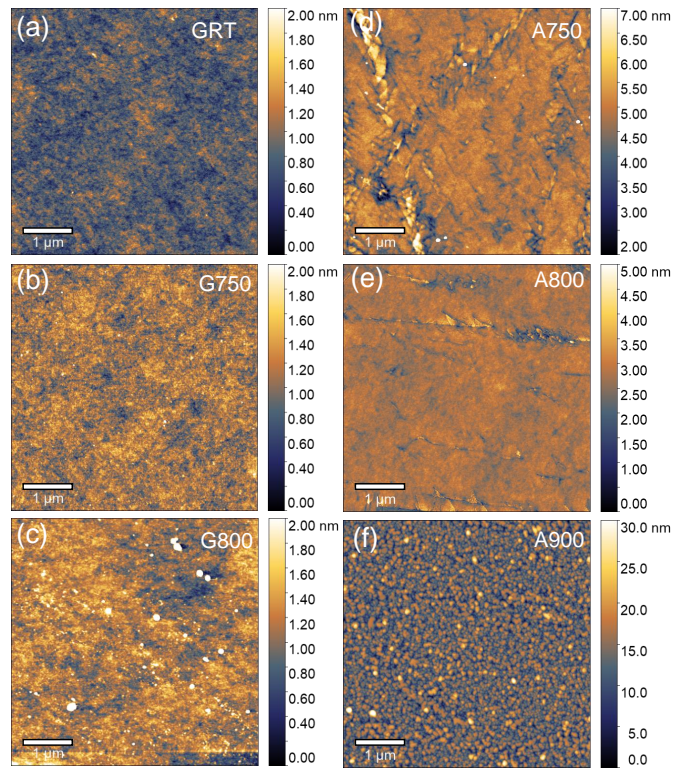

FIG. 2. Surface quality examination by AFM. 5μm x 5μm AFM topography images of 22-29 nm thick TbIG films samples: (a) as-grown at RT, (b) grown at 750 °C, (c) grown at 800 °C, grown at RT and annealed at (d) 750 °C, (e) 800 °C and (f) 900 °C.

Figure 2 shows the AFM topography maps of selected samples to assess the quality and roughness of the surfaces. While the surface of the GRT films is smooth and flat (r.m.s. roughness 130 pm), upon increasing temperature, the number of defects and the overall roughness increase. After post-annealing crack-like features appear [Fig. 2(d)-(e)], affecting the overall roughness of the surface, which are not present in the samples grown at high temperatures. The sample annealed at 900 °C (A900) also displays a granular morphology, observed previously in other REIGs,[32] which is detrimental for interfacial spin transport when a metallic layer is deposited on the film for device fabrication. The increased roughness is also reflected in the decrease of the amplitude of the XRR oscilla-tions that are almost absent for sample A900. However, in general the amplitude of the XRR oscillations for all samples (see Fig. 1(b),(d),(f)) is lower than what expected for the observed surface roughness below 1 nm. This could indicate that the interface between the film and the substrate is not sharp, but some intermixing of the atoms occurs.

TABLE I. Summary of the deposition and various parameters.

| Sample | Sub. $a$(Å) | Growth T (°C) | Anneal. T (°C) | T-S (mm) | $d_{444}$ (Å$^{-1}$) | Strain (%) | FWHM (°) | Thick. (nm) | Tb:Fe | Rough. (rms) |
|---|---|---|---|---|---|---|---|---|---|---|
| GRT | 12.38 | RT | none | 150 | no | - | - | 29 | - | 130 pm |
| A700 | 12.38 | RT | 700 | 150 | no | - | - | 26 | - | 140 pm |
| G750 | 12.38 | 750 | none | 150 | 1.8440 | 2.535 | 0.323 | 22 | 0.51 | 150 pm |
| A750 | 12.38 | RT | 750 | 150 | 1.8173 | 1.051 | 0.372 | 25 | - | 250 pm |
| G800 | 12.38 | 800 | none | 150 | 1.8411 | 2.374 | 0.336 | 22 | 0.52 | 150 pm |
| A800 | 12.38 | RT | 800 | 150 | 1.8187 | 1.219 | 0.403 | 25 | - | 200 pm |
| A900 | 12.38 | RT | 900 | 150 | 008149 | 0.917 | 0.514 | - | - | 2 nm |
| G800 +8 | 12.38 | 800 | none | 158 | 1.8474 | 2.725 | 0.334 | 22 | 0.39 | - |
| G800 -8 | 12.38 | 800 | none | 142 | 1.8400 | 2.313 | 0.414 | 23 | 0.55 | - |
| G800 -12 | 12.38 | 800 | none | 138 | 1.8372 | 2.157 | 0.415 | 21 | 0.59 | - |
| G750 S48 | 12.48 | 750 | none | 150 | 1.8259 | 1.529 | 0.331 | 25 | 0.58 | - |
| G750 S50 | 12.50 | 750 | none | 150 | 1.8322 | 1.879 | 0.318 | 22 | 0.66 | - |

We have also tested the effect of oxygen partial pressure used during growth on the surface quality (images in the Supplementary Material). All the samples discussed here were prepared with 6% $O_2$, which yielded the smoother surface among the parameters tested. Due to the better crystal and surface quality of the films grown directly on a heated substrate, the rest of this study focuses on such samples.

Compositional analysis of the TbIG films was carried out using x-ray photoelectron spectroscopy (XPS). Figure 3 shows representative survey and high-resolution XPS spectra of the samples. In all cases, the observed peaks can be attributed to Fe and Tb in the 3+ oxidation state; thus, if present, the amount of other species or defects is beyond the resolution of the technique. From the Fe $2p$ and Tb $3d$ regions of the XPS spectra, the Tb:Fe ratio is also extracted (see Supplementary Material for details). For the majority of the samples the Fe:Tb ratio is below the bulk value of 0.6. This indicates a Tb deficiency in the films, which is opposite to what was previously observed in PLD-grown samples that were found to be RE-rich.[21,33] This deviation from the bulk stoichiometry has relevant consequences on the sample's magnetic properties, especially $T_M$.

Systematically reducing the T-S distance during growth leads to an increase in the Tb content in the films (see Table I). This can be explained by a loss of Tb caused by the scattering of the larger Tb atoms with the reactive gas in the sputtering chamber. Specifically, the Tb content can be increased from 0.39 to 0.59 by reducing the T-S distance from 158 to 138 mm. The composition of the samples grown on SGGG (111) was also measured by XPS. In this case, the Tb:Fe ratio is closer to the bulk value. A plausible explanation for this difference is the formation of the Fe vacancies in the Garnet lattice to accommodate the tensile strain from the SGGG substrates. The result is a smaller lattice parameter of the film compared to the samples grown on GGG (see Fig. 1(e)), despite the larger lattice constant of the substrate, and hence larger Tb:Fe ratio. The presence of more crystal defects in these samples is also reflected in the magnetic properties, namely lower $H_c$ and amplitude of the AHE, as discussed below.

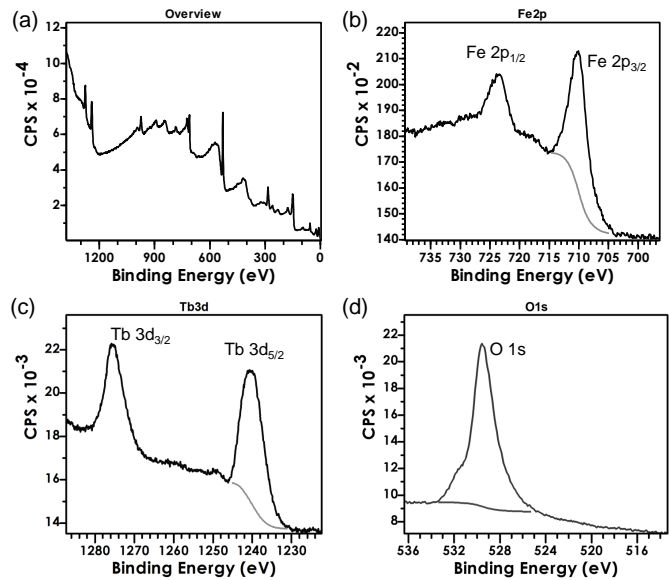

FIG. 3. XPS analysis. (a) Survey scan (b) Fe $2p$, (c) Tb $3d$ and (d) O $1s$ high-resolution spectra of a 22 nm-thick TbIG film grown at 800 °C on GGG (111).

### B. Magnetic properties

The crystalline TbIG samples of this study are ferrimagnetic and display square magnetic hysteresis loops at room temperature, which indicates PMA. A representative out-of-plane hysteresis loop of the sample G800 measured by polar MOKE is shown in Fig. 4(a). The hysteresis loops were also measured by SQUID-VSM, and a representative plot is shown in Fig. 4(b). Here, the out-of-plane magnetization displays sharp switching and square hysteresis loop with some additional bumps unrelated to the TbIG magnetization and likely due to nonlinear background from the substrate. The low saturation value of approximately 50 emu cm$^{-3}$ is expected for a ferrimagnet close to $T_M$. On the other hand, the in-plane $M$ only reaches 20 emu cm$^{-3}$ in the same field range used for the out-of-plane measurement, indicating strong PMA.

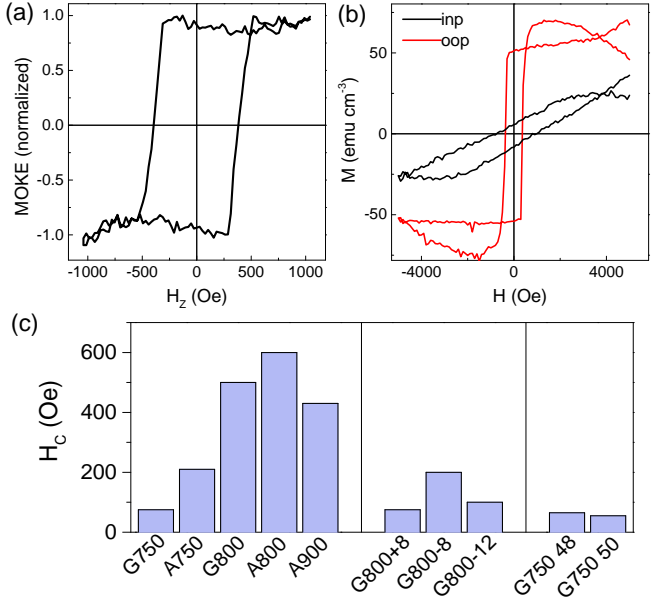

FIG. 4. Room temperature magnetic properties of a 25 nm thick TbIG film grown at 800 °C on GGG (111). a) Out-of-plane MOKE signal. b) Magnetization ($M$) as a function of in-plane (black) and out-of-plane (red) field ($H$) measured by SQUID. c) Comparison of coercivity values upon varying deposition conditions: annealing/growth temperature, T-S distance and substrate.

Performing MOKE measurements at RT on samples prepared in different thermal conditions, on different substrates, and with different T-S distances reveal systematic changes to $H_c$, as summarized in Fig. 4(c). With the increasing annealing or growth temperature, we observe a significant increase in $H_c$. We also find that the post-annealed samples have a slightly larger $H_c$ than the ones directly grown at the same temperature. The only exception to this trend is for the samples annealed ex-situ at 900 °C (A900). The smaller $H_c$ in this film compared to A800 is attributed to a lower PMA caused by a lower strain and poorer surface quality revealed by the XRD and AFM studies, respectively (see Fig. 1(a)-(b)).

On the other hand, when the T-S distance is lowered from 150 to 138 mm a drastic reduction of the RT coercivity is observed. Such a decrease in $H_c$ can be explained by the reduction in compressive strain with the decreasing T-S distance (see Fig. 1(c)), leading to a weaker PMA. Surprisingly, a higher T-S distance (158 mm) also shows a drop in $H_c$, despite the larger compressive strain. This can be tentatively explained by the reduction in Tb content causing a decrease in $T_M$, thus pushing $T_M$ further away from RT, which could cause a decrease in PMA and $H_c$ at RT. The temperature-dependent coercivity is further investigated by measuring the AHE at lower temperatures, which is the subject of the next section.

Finally, The films grown on SGGG also display PMA with square out-of-plane hysteresis and room-temperature coercivity below 100 Oe. This is consistent with the lower compressive strain (see Fig. 1(e)) compared to the films deposited on GGG substrates.

## C. Interfacial spin transport properties

Interfacial spin transport properties of Pt/TbIG heterostructures were investigated through AHE measurements. The AHE in this bilayer system predominantly occurs due to spin Hall magnetoresistance proportional to the imaginary part of the spin mixing conductance of the Pt/TbIG interface.[34] In principle, the AHE alone cannot be a quantitative measure of the interfacial spin transport in such films since its amplitude sensitively depends on the spin Hall angle, thickness, and magnetic proximity effects of Pt, interfacial spin memory loss, and other interface and bulk-related properties.[35] However, similar surface morphology in the optimized TbIG films revealed by AFM topography images and identical Pt deposition conditions ensure that most of these contributions would be comparable in our films. Therefore, we tentatively use AHE as a figure of merit for the interfacial spin transparency in this study for qualitatively comparing TbIG films prepared in different conditions.

To detect the AHE, we measure the first harmonic Hall resistance ($R_H$), which contains the following contributions:[36]

$$R_\omega = R_H = R_{PHE} \sin 2\theta \sin 2\phi + R_{AHE} \cos\theta + R_{OHE} H \cos\theta \quad (1)$$

where $R_{PHE}$ and $R_{AHE}$ represent the transverse SMR-induced planar and anomalous Hall resistances, respectively, and $R_{OHE}$ is the ordinary Hall resistance of Pt. The polar coordinate system and experimental geometry are represented in Fig. 5(a): $H_z$ is the out-of-plane (z) component of the external field while $\phi$ and $\theta$ represent the magnetization angles with the $x$ and $z$ axes, respectively. To measure the $R_{AHE}$ contribution given in Eq.(1), we measure the transverse voltage ($V_H$) while sweeping the external field along z ($\theta=0°$).

Figure 5(b) shows a representative plot of the $R_{AHE}$ vs. $H_Z$ recorded at RT in Pt grown on TbIG G800 sample (see growth details in Section II). The square shape of the hysteresis loops confirms the presence of PMA, and the magnitude of $H_c$ is in excellent agreement with the MOKE measurements on the continuous film before patterning, indicating that the film did not suffer any changes during lithography and Pt deposition.

The amplitude of the AHE at RT is the largest (1.7 mΩ) within all samples in this study, and is comparable with previous measurements on TbIG/Pt films where TbIG was grown by PLD.[21,31] Reducing the growth temperature, varying the T-S distance during growth, or growing the films on SGGG result in a reduction of the amplitude of the AHE [Fig. 5(c)]. Therefore we conclude that the growth parameters yielding the best structural and surface properties (sample G800) also give rise to the highest AHE and hence interfacial spin transparency.

## D. Magnetic compensation temperature

In ferrimagnetic materials, particularly in REIGs, the magnetization of the individual sublattices evolves differently with temperature. As customary in REIGs, the magnetization of TbIG is dictated by two antiparallel magnetic sublattices, the first one comprised of three tetrahedral $Fe^{3+}$ ions and the second comprised of two octahedral $Fe^{3+}$ and three dodecahedral magnetic $RE^{3+}$ ions.[37] Due to the strong temperature dependence of the Tb magnetic moment, bulk TbIG shows magnetic compensation (i.e., $M_{net} = 0$) at $T_M = 248$ K. Approaching this temperature, the vanishingly small Zeeman energy makes it increasingly more difficult to switch the magnetization of TbIG between the two stable states, and hence $H_c$ shows diverging behavior. In TbIG, for $T < T_M$ the net magnetization is expected to align with the direction of the Tb-containing sublattice, while for $T > T_M$, it mainly follows the tetrahedral $Fe^{3+}$ sublattice. For ultrathin films, due to the strong paramagnetic background from the GGG substrate, it is generally difficult to detect $T_M$ by standard magnetometry. There is an additional difficulty in measuring $T_M$ if this value is not strictly constant throughout the film surface due to varying stoichiometry, thickness, etc. Recently, it has been reported that the sign of the AHE in REIG/Pt changes when crossing $T_M$.[31,38] This simple method combined with diverging $H_c$ behavior thus offers an ideal platform to detect $T_M$ when AHE is finite.

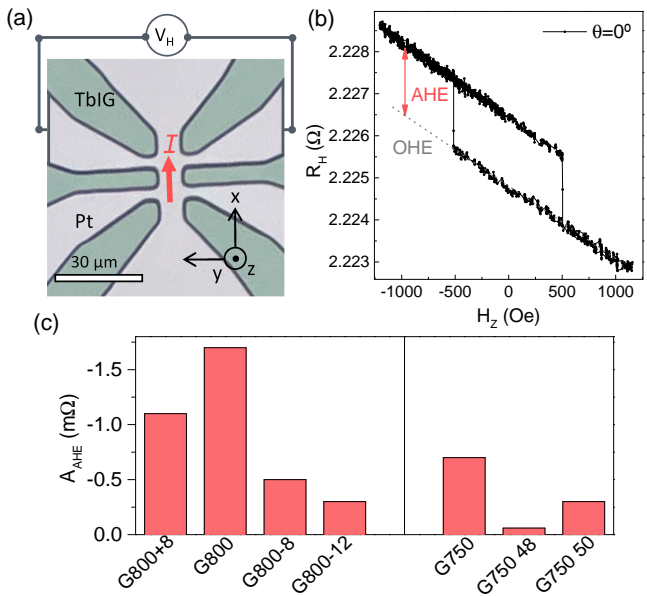

FIG. 5. Room temperature magnetotransport properties of Pt/TbIG bilayers when TbIG was grown at 800 °C on GGG (111). a) Optical microscopy image of the Hall bar device and the measurement scheme and (b) plot of the Hall resistance ($R_H$) as a function of out-of-plane field ($H_Z$). c) Comparison of amplitude of the AHE ($A_{AHE}$) upon varying T-S distance and GGG substrate.

Figure 6(a) shows the temperature evolution of the AHE for Pt/TbIG bilayer when TbIG was grown at 750 °C on the standard GGG substrate. As evident from the data, the AHE is negative at 220 K and $H_c$ increases rapidly with decreasing the temperature until $T$=190 K, where the coercivity exceeds 1 T. Below this temperature, the AHE changes sign, and $H_c$ starts to decrease with the decreasing temperature. Based on these measurements, we determine $T_M$=190 K, well below the bulk value of 248 K. Further decreasing the temperature, the amplitude of the AHE decreases monotonically and vanishes at $T_1$=150 K. This temperature does not correspond to an anomaly in the coercive field, as shown in Fig. 6(b), indicating that the disappearance of the AHE is not related to the ferrimagnetism in TbIG. A similar sign change of AHE at low temperatures has been observed in other systems, such as Pt/YIG[39] and W/TmIG,[40] and attributed to magnetic proximity effects. The $T$ dependence of the amplitude of the AHE and its absolute value are summarized in Fig. 6(c).

To investigate the role of film stoichiometry on $T_M$, we repeat similar temperature-dependent AHE measurements for the series of TbIG samples grown with different T-S distances. We remind that the XPS data suggest that our films are generally Tb-deficient, and a systematic variation of Tb:Fe ratio can be obtained depending on the T-S distance. Closer T-S leads to better stoichiometry, which could potentially increase $T_M$. Figure 7 shows the evolution of $H_c$ as a function of temperature for four different T-S distances. In all films, the estimated $T_M$ is between 190 K and 225 K with no systematic variation with respect to T-S distance and hence Tb:Fe ratio. Therefore, although the Tb:Fe ratio can be brought close to bulk (see Table I), no significant increase in $T_M$ is observed.

Assuming that the XPS findings of the Tb:Fe ratios are generally accurate (at least the qualitative trend with the T-S distance), the temperature-dependent AHE data suggest that the stoichiometry alone cannot be the only parameter to modify $T_M$. Indeed, Rosenberg et al.[21] and Ren et al.[41] found a higher $T_M$ of about 350 K in Fe-deficient TbIG films, hence interpreted their results in view of stoichiometry. On the other hand, Liu et al.[31] reported a $T_M$ close to 220 K and attributed to strain effect or oxygen vacancies. Therefore, there is still no systematic understanding of the $T_M$ in ultrathin REIG films. Since, in our case, stoichiometry does not seem to be the dominant factor, an important role could be played by the nature of the vacancies in the garnet lattice. It is crucial to know if the empty Tb sites are replaced by Fe atoms or not, and if so, with which oxidation state. These fine details will have a significant effect on the net sublattice magnetization and, thus, on the compensation temperature. Further experiments are needed to investigate the nature of the Tb deficiency in our films from a crystallographic point of view.

## IV. CONCLUSIONS

In summary, we report the growth and extensive characterization of high-quality TbIG films on GGG (111) and SGGG (111) substrates by RF magnetron sputtering. The films exhibit PMA, induced by compressive strain and tunable properties through annealing conditions, T-S distance and substrate choice. In particular, growing the films directly on heated substrates results in better crystallinity, larger epitaxial strain, larger $M_S$ and lower $H_c$ compared to films grown at room temperature and annealed subsequently. The AFM topography images show that the optimized films have roughness well below 1 nm. The XPS characterization of the films reveals that the films are, in general, Tb-deficient due to the higher prob-ability of Tb scattering with Ar plasma. The Tb deficiency can be partially corrected by growing the films at a lower T-S distance at a cost of reduced strain and hence PMA. The AHE measurements in Pt/TbIG bilayer indicate good interface spin transparency, especially when using structurally and magnetically optimized TbIG films for such devices. The sign change of the AHE and diverging $H_c$ are employed to examine the $T_M$ of selected films. We find that all examined films, independently of the Tb:Fe ratio, exhibit $T_M$ values near 200 K, some 50 K lower than in bulk TbIG. Our results indicate that the compensation temperature in TbIG does not dependent in a simple way on modifications in the Tb:Fe ratio, but could be affected by other factors such as the nature of the Tb vacancies in the garnet lattice.

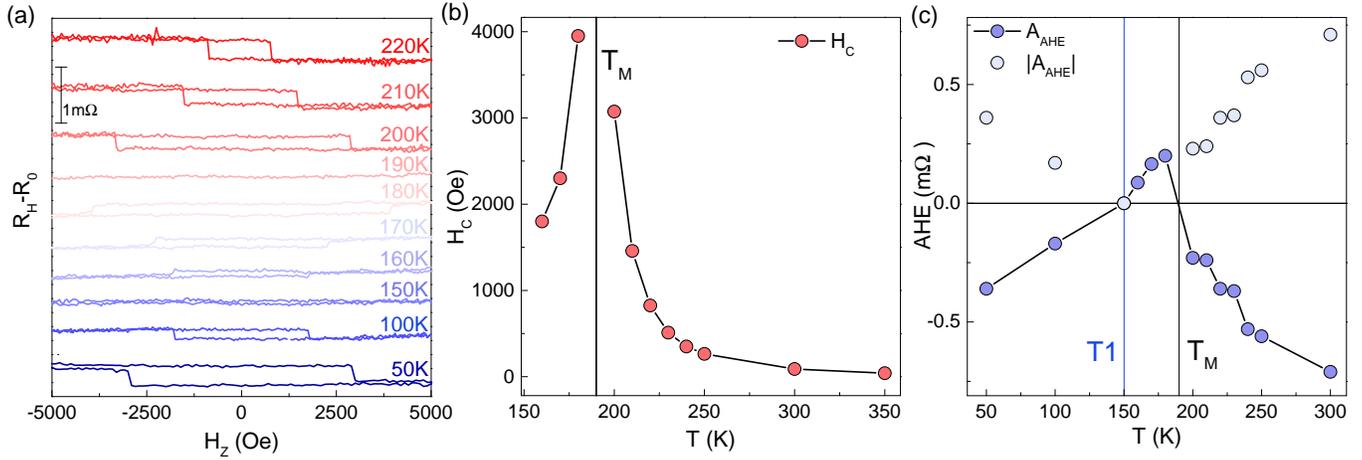

FIG. 6. Temperature dependence of the AHE of Pt/TbIG when TbIG was grown at 750 °C. a) Hall resistance ($R_H$) as a function of out-of-plane field ($H_z$) measured between 160 and 220 K. The linear contribution of the OHE in Pt has been subtracted from the data. b) Coercive field ($H_c$) and (c) amplitude of the AHE ($R_{AHE}$) as a function of temperature ($T$) extracted from panel (a). The compensation temperature ($T_M$=190 K) and the temperature at which the AHE is zero ($T_1$=150 K) are indicated with black and blue lines, respectively.

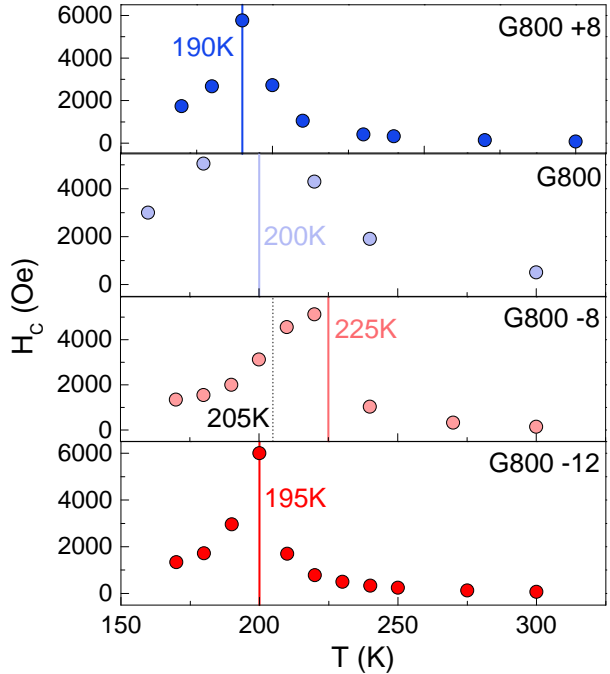

FIG. 7. Temperature ($T$) dependence of the coercive field ($H_c$) of Pt/TbIG grown at 800 °C with varying T-S from 158(+8) mm (top) to 138(-12) mm (bottom). The black dotted line indicates to the sign-reversal of the AHE that for the G800 -8 sample does not correspond to the maximum of $H_c$.

## SUPPLEMENTARY MATERIAL

See Supplementary Material for further details of the XPS data analysis, RSM maps, AFM images and the full set of MOKE and Hall effect data not shown in the main manuscript.

## ACKNOWLEDGMENTS

We would like to thank Prof. Josep Funtcuberta, Dr. Martín Testa Anta and Stefano Fedel for the insightful discussions. The authors acknowledge funding from the European Research Council (ERC) under the European Union's Horizon 2020 research and innovation programme (project MAG-NEPIC, grant agreement No. 949052).

## AUTHOR DECLARATIONS

### Conflict of Interest

The authors have no conflicts to disclose

**Author Contributions**

**Silvia Damerio**: Conceptualization – Ideas (equal); Data Curation (lead); Formal Analysis (lead); Investigation (equal); Methodology (equal); Visualization (lead); Writing/Original Draft Preparation (lead); Writing/Review and Editing (equal).

**Can Onur Avci**: Conceptualization – Ideas (equal); Data Curation (supporting); Formal Analysis (supporting); Funding Acquisition (lead); Investigation (equal); Methodology (equal); Supervision (lead); Visualization (supporting); Writing/Original Draft Preparation (supporting); Writing/Review and Editing (lead).

## DATA AVAILABILITY

The data that support the findings of this study are available from the corresponding author upon reasonable request.

# Sputtered terbium iron garnet films with perpendicular magnetic anisotropy for spintronic applications

Silvia Damerio, Can Onur Avci

*Institut de Ciencia de Materials de Barcelona (ICMAB-CSIC), Campus de la UAB, 08193 Bellaterra, Spain*

### S1) XPS Quantification of composition

X-Ray photoelectron spectra (XPS) of TbIG samples were collected on a SPECS PHOIBOS 150 system and analyzed by the software CasaXPS. To calculate the Tb:Fe ratio, the area of the Tb $3d_{5/2}$ and Fe $2p_{3/2}$ peaks was measured and normalized by the relative scattering factor (RSF) in the software element library, 49.42 and 10.82 respectively. The accuracy of quantification of the XPS technique is usually ~10%, however in our case further uncertainty is added by the isolating nature of the GGG substrate that can cause some charging effects that are reflected in the spectra in the form of a shoulder to the low energy side of the XPS peaks. Therefore, to take into account the uncertainty introduced by this artifact, we calculated the area of the peak both with and without shoulder, as shown Fig. S1.1, and took the average between the two. Fig. S1.2 shows the Tb:Fe concentration as a function of T-S distance. The error bars are given by the difference between the quantification with and without peak shoulder. Although the absolute value of the Tb:Fe has a large uncertainty, the trend is clear: the amount of Tb increases as the substrate goes closer to the target.

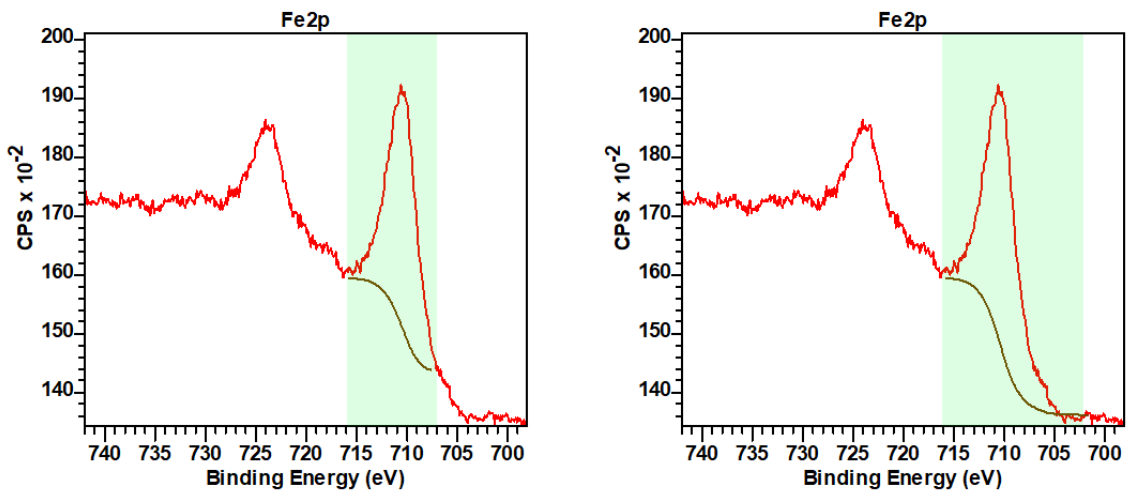

*Fig. S1.1 Fe 2p region of the XPS spectrum of a TbIG sample showing the Fe $2p_{3/2}$ peak area without (a) and (b) with shoulder of the peak.*

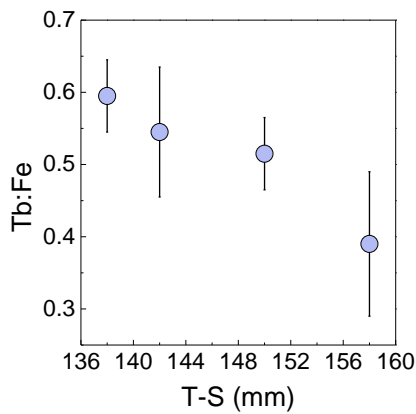

*Fig. S1.2 Plot of the Tb:Fe ratio as a function of target-substrate distance.*

## S2) RT MOKE

The magnetic hysteresis loops measured for all the samples of the paper by polar MOKE are shown in Fig. S2.

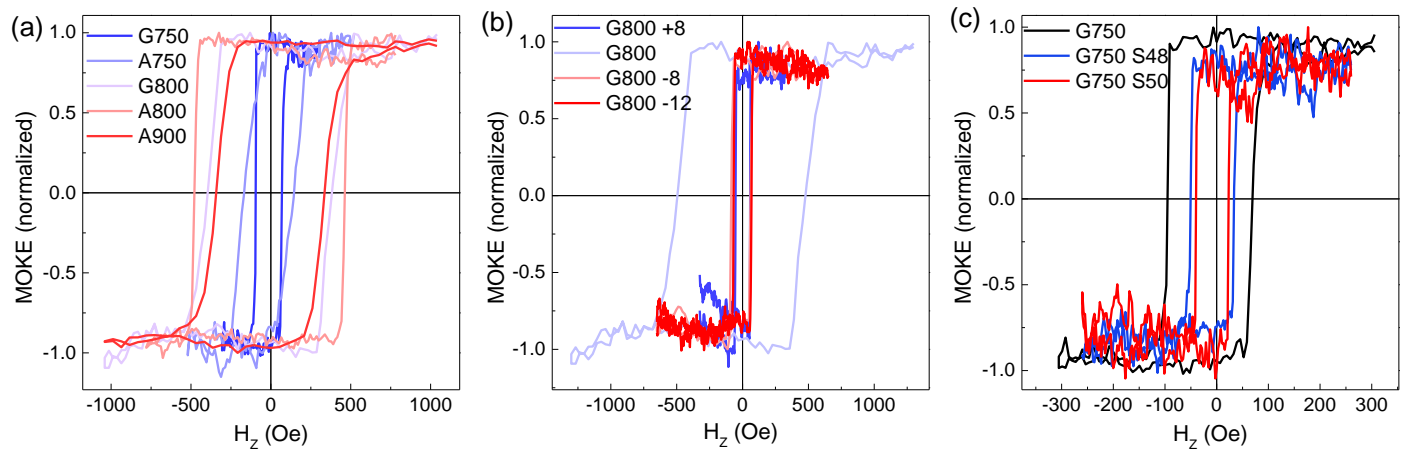

*Fig.S2 Plot of the MOKE signal as a function of out-of-plane field ($H_Z$) for samples (a) grown/annealed at different temperatures, (b) grown at different T-S distance and (c) on different GGG substrates.*

## S3) RT Hall Resistance

The room temperature AHE was measured for all the samples described in the paper. Fig. S3 shows the plots of $R_H$ as a function of $H_Z$ summarized in Fig. 4(c) and 5(c) of the main paper.

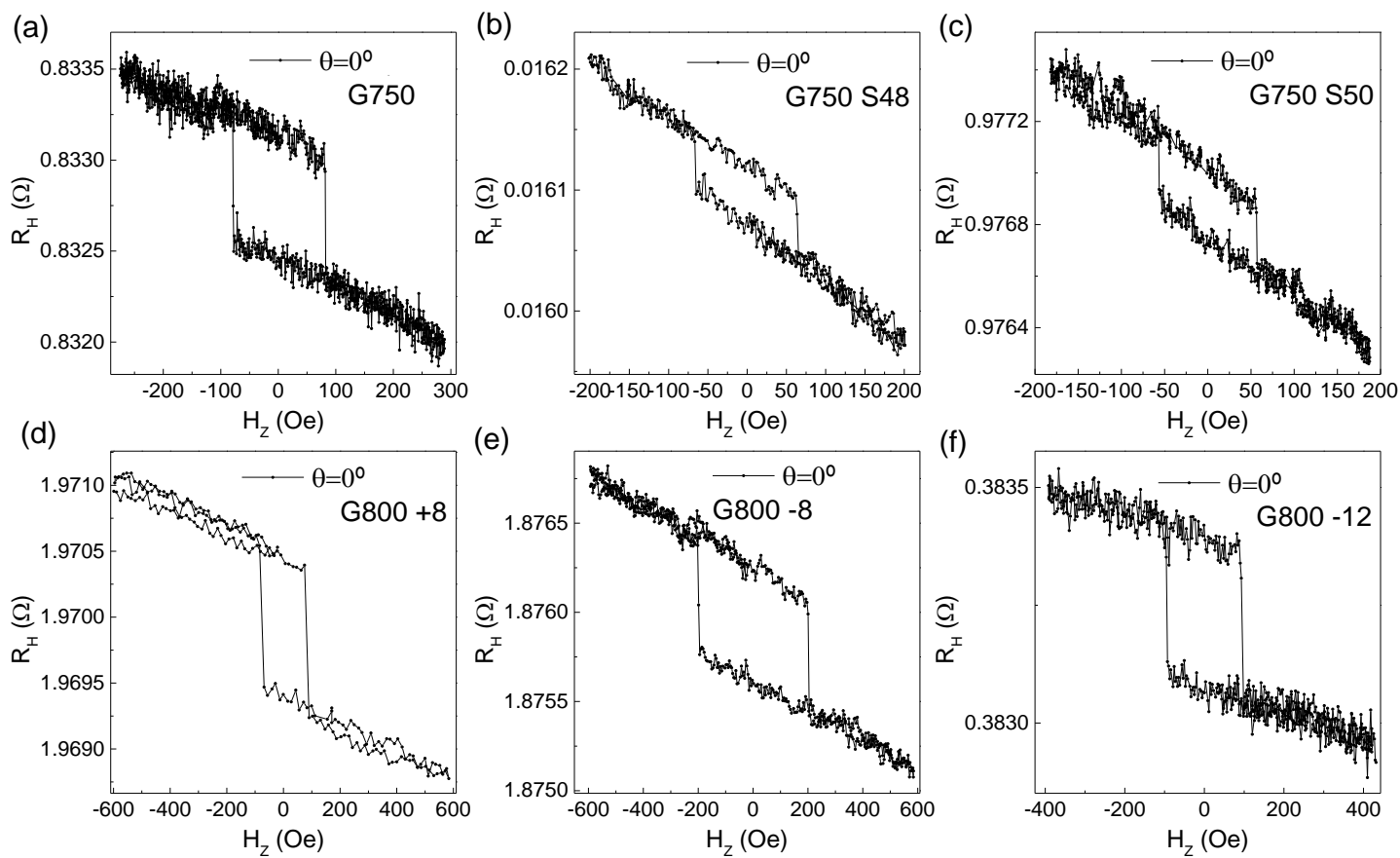

*Fig.S3 Plot of the first Harmonic Hall resistance ($R_H$) as a functions od out-of-plane field ($H_Z$) of TbIG films grown: (a) at 750 ºC on GGG and T-S distance 150 mm, (b) at 750 ºC on SGGG-48 and T-S distance 150 mm, (c) at 750 ºC on SGGG-50 and T-S distance 150 mm, ( d) at 800 ºC on GGG and T-S distance 158 (+8) mm , (e) at 800 ºC on GGG and T-S distance 142 (-8) mm and (f) at 800 ºC on GGG and T-S distance 138 (-12) mm.*

## S4) Temperature dependence of the anomalous Hall effect

Fig. S4 shows the plot of $R_H$ as a function of field measured at 5 selected temperatures for samples grown at variable T-S distance. A graphical summary of the data is depicted in Fig. 6 of the main text.

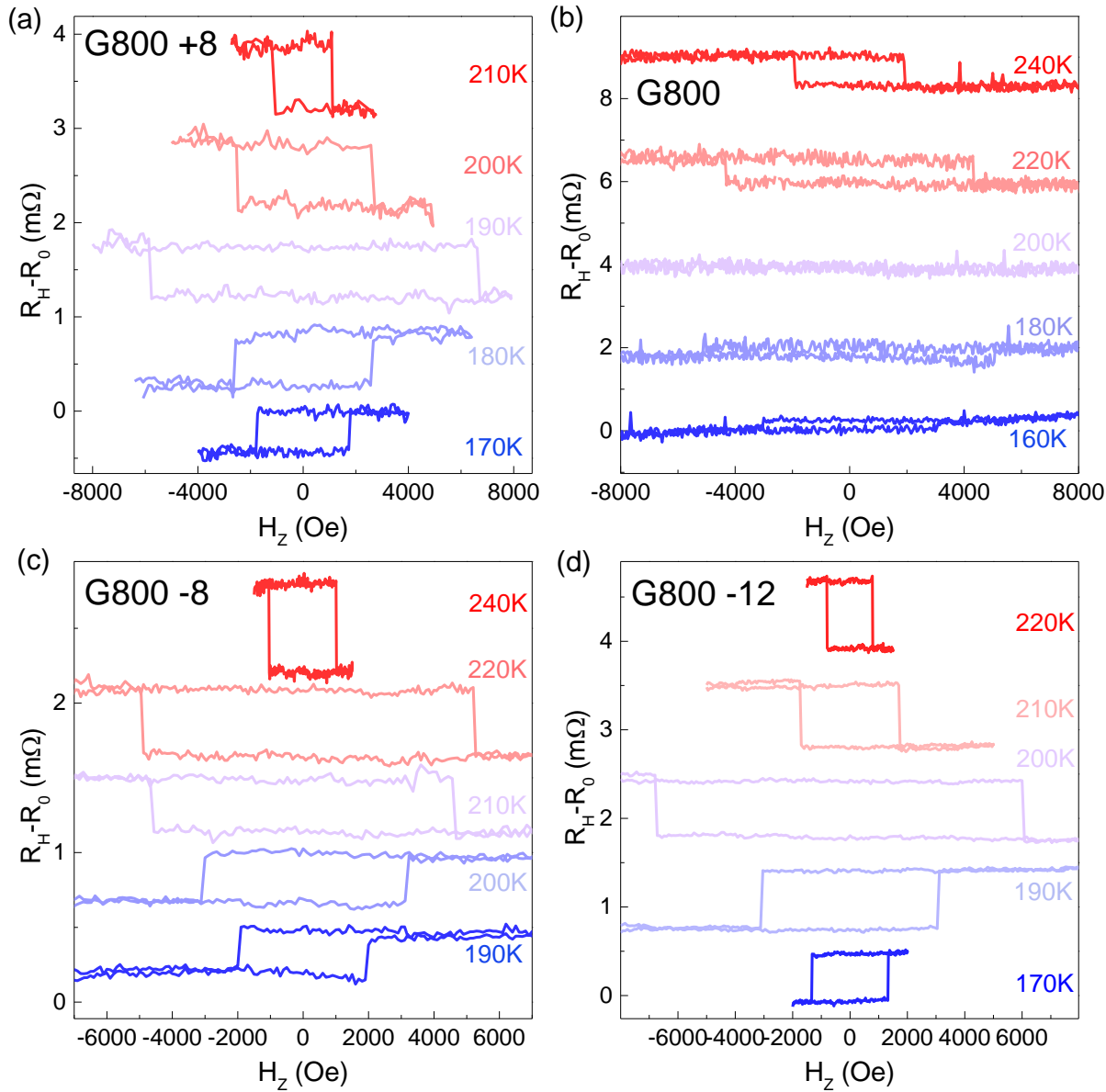

*Fig.S4 Plot of the hall resistivity ($R_H$) as a function of out-of-plane field measured at 5 selected temperatures close to compensation for samples grown at 800 ºC on GGG(111) at varying T-S distance: (a) 158 (+8) mm, (b) 150 mm, (d) 142 (-8) mm and (e) 138 (+8) mm.*

## S5) Reciprocal space maps

Figure S5 shows the reciprocal space maps of the (642) reflection of samples G750, G800, G750 S48 and G750 S50. As it can be seen, in all cases the film and substrate peaks are vertically aligned, indicating that the films are fully strained to the substrate. The in-plane strain can therefore be calculated as: $\frac{d_{1-10}-d_B}{d_B} = \frac{d_S-d_B}{d_B}$, where $d_B$ is the bulk lattice parameter and $d_{1-10}$ is the in-plane lattice parameter of the film that corresponds to the one of the substrate ($d_S$). For the samples grown on GGG this results in a compressive strain of 0.4190 %, while for the samples grown on SGGG(48) and SGGG(50) in a tensile strain of -0.3084% and -0.5358% respectively. In the presence of compressive strain, the unit cell is elongated in the out-of-plane direction, and thus the film peak appears at lower $q_z$ values than the substrate [Fig. S5(a)-(b)]. On the other hand, in the presence of tensile strain, the out-of-plane contraction of the unit cell should lead to a smaller lattice parameter and thus the film peak should appear at a larger $q_z$ value than the substrate. However, this is not what we find for the films grown on SGGG [Fig. S5(c)-(d)], where the film peak is closer to that of the substrate, but still at a lower $q_z$. This is consistent with the out-of-plane peak shift observed in the 2θ-ω scans of Fig.1 of the main text and the derived values of out-of-plane strain listed in Table 1 of the main text. Therefore, we deduce that the samples deposited on SGGG are also compressively strained, which explains the presence of PMA.

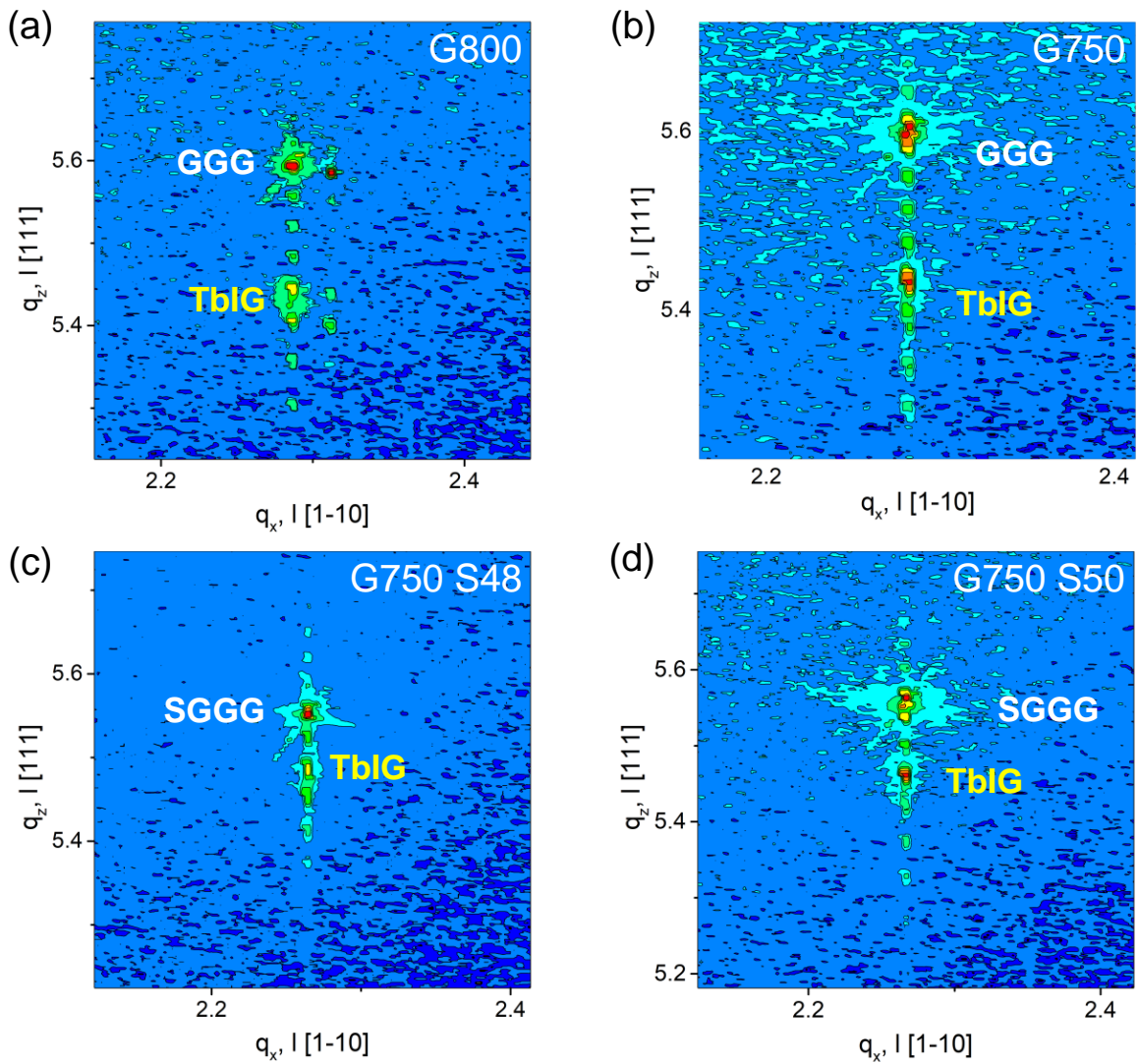

*Fig.S5 Reciprocal space maps of the (642) reflection of 22-25 nm thick TbIG films grown on GGG and SGGG substrates. (a-b) Samples grown on GGG (111) at 800 ºC and 750 ºC respectively. (c-d) Samples grown on SGGG (111) with lattice constant a=12.48 and a=12.50.*

## S6) Effect of O₂ partial pressure on surface quality

In general, for sputtering deposition, the $O_2$:Ar ratio is kept low in order to maintain the plasma and obtain a relatively fast deposition rate. In this work we used a $O_2$:Ar ratio of 2:30 with 3 mtorr total pressure. Therefore, we expect that small modifications in the $O_2$ partial pressure will not play a large role in tuning the films stoichiometry compared to other techniques, such as PLD, where $O_2$ is the only process gas. However, during the sputtering deposition process a built-in electric field is formed that accelerate the $Ar^+$ cations towards the target and the $O^{2-}$ anions towards the substrate. Bombardment by such ions could influence the quality of the crystal structure and roughness of the surface. Therefore, to investigate this aspect, we have deposited a new set of samples varying $O_2$ partial pressure during growth meanwhile maintaining all the other parameters constant (substrate: GGG (111), temperature: 800 ºC, power: 150 W, deposition time: 3600 s, T-S: 150 mm, Ar flow rate 30 sccm, cooling: -0.5 º/s in 50 mtorr $O_2$). The $O_2$ flow rates chosen were 0 sccm (0%), 1 sccm (3%), 2 sccm (reference 6%) and 4 sccm (12%). Figure 3 shows the AFM images of the samples' surface. As it can be seen, smoother surfaces are obtained when the $O_2$ flow rate is 1 or 2 sccm [Fig. R3(b)-(c)], with rms roughness of about 120 pm, as in Fig. 2 of the main text. When no oxygen is added during growth [Fig. R3(a)] the morphology of the surface changes slightly and a grain-like texture can be seen that increases the rms roughness (200 pm). This could be explained by a different crystallization of the TbIG in pure Ar atmosphere. Finally, at a higher $O_2$ flow rate of 4 sccm [Fig. R3(d)] some crack-like features similar to those observed in the post-annealed samples appear that are also detrimental to the overall surface quality (roughness rms 170 pm). Therefore, we conclude that some oxygen content during growth is necessary for optimum crystallization of TbIG, but as the referee pointed out, tuning the oxygen percentage during growth has an influence on the surface quality. In our case oxygen percentages between 3 and 6 % seem to be best from this point of view.

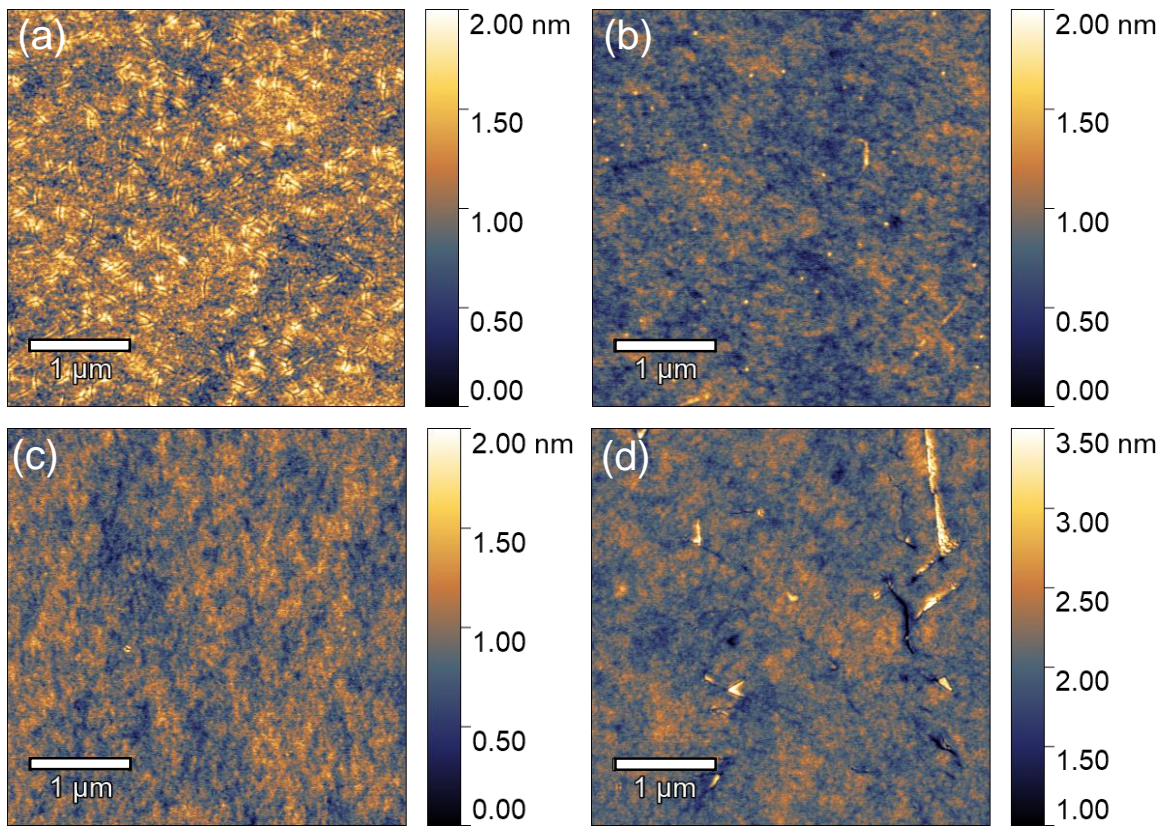

Fig.S6 AFM topography images of TbIG films deposited with varying $O_2$ flow rate: (a): 0 sccm, (b) 1 sccm, (c) 2 sccm and (d) 3 sccm.